\pdfoutput=1

\documentclass{iau}
\usepackage{graphicx}

\newcommand\chandra{\textit{Chandra}}
\newcommand\dataset[2][junk]{#2}
\newcommand\pram{p_{\rm hotspot}}

\title[Cygnus A and its environment] 
{Interaction of Cygnus A with its environment}

\author[P.~Nulsen et al.]   
{
Paul E. J. Nulsen$^1$,
Andrew J. Young$^2$,
Ralph P. Kraft$^1$,
Brian~R.~McNamara$^3$
\and
Michael W. Wise$^4$
}

\affiliation{
$^1$Harvard-Smithsonian Center for Astrophysics, \\ 
60 Garden Street, Cambridge, MA 02138, USA \\ 
email: {\tt pnulsen@cfa.harvard.edu, rkraft@cfa.harvard.edu} \\
[\affilskip]
$^2$School of Physics, University of Bristol, \\ 
Tyndall Avenue, Bristol BS8 1TL, UK \\
email: {\tt andy.young@bristol.ac.uk} \\
[\affilskip]
$^3$Department of Physics and Astronomy, University of Waterloo, \\
200 University Avenue West, Waterloo, Ontario, Canada N2L 3G1 \\
email: {\tt mcnamara@uwaterloo.ca} \\
[\affilskip]
$^4$ASTRON, Netherlands Institute for Radio Astronomy \\
P.O. Box 2, 7990 AA Dwingeloo, Netherlands \\
email: {\tt wise@astron.nl}
}

\pubyear{2014}
\volume{313}
\pagerange{xx}
\jname{Extragalactic jets from every angle}
\editors{F. Massaro, C.C. Cheung, E. Lopez, A. Siemiginowska, eds.}

\begin{document}

\maketitle

\begin{abstract}
Cygnus A, the nearest truly powerful radio galaxy, resides at the
centre of a massive galaxy cluster.  \chandra{} X-ray observations
reveal its cocoon shocks, radio lobe cavities and an X-ray jet, which
are discussed here.  It is argued that X-ray emission from the outer
regions of the cocoon shocks is nonthermal.  The X-ray jets are best
interpreted as synchrotron emission, suggesting that they, rather than
the radio jets, are the path of energy flow from the nucleus to the
hotspots.  In that case, a model shows that the jet flow is
non-relativistic and carries in excess of one solar mass per year. 
\end{abstract}

\firstsection 
\section{Introduction}

At a redshift of 0.056, the Fanaroff-Riley class II radio galaxy
Cygnus A is the nearest of the truly powerful radio galaxies, much
closer than any comparable sources (\cite{cb96}).  As a result, it the
archetype of powerful radio galaxies.  The galaxy that hosts Cygnus A
is also the central galaxy of a massive galaxy cluster, so that
interactions between the expanding radio lobes of Cygnus A and the
surrounding gas can be observed in the X-ray (\eg \cite{cph94},
\cite{swa02}).  Here we discuss some properties of Cygnus A determined
from \chandra{} X-ray observations.

\section{The lobes and shocks}

\begin{figure}[t]
\begin{center}
\includegraphics[width=0.78\textwidth]{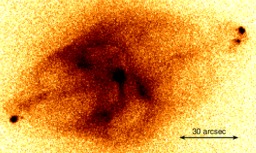} 
\includegraphics[width=0.78\textwidth]{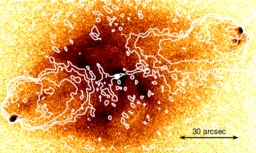}
 \caption{Background subtracted, exposure corrected 0.5 -- 7 keV
   \chandra{} image of Cygnus A.  The lower image also shows contours
   of the 6 cm radio emission from the map of \cite{pdc84}; contour
   levels 0.001, 0.003, 0.01, 0.03, 0.1 $\rm Jy\
   beam^{-1}$).} \label{fig:cyga}  
\end{center}
\end{figure}

\dataset[ADS/Sa.CXO#obs/00360]{}
\dataset[ADS/Sa.CXO#obs/01707]{}
\dataset[ADS/Sa.CXO#obs/05830]{}
\dataset[ADS/Sa.CXO#obs/05831]{}
\dataset[ADS/Sa.CXO#obs/06225]{}
\dataset[ADS/Sa.CXO#obs/06226]{}
\dataset[ADS/Sa.CXO#obs/06228]{}
\dataset[ADS/Sa.CXO#obs/06229]{}
\dataset[ADS/Sa.CXO#obs/06250]{}
\dataset[ADS/Sa.CXO#obs/06252]{}

The 0.5 -- 7 keV image of Cygnus A in Fig.~\ref{fig:cyga} was made
from 246 ksec of cleaned \chandra{} data obtained between 2000 and
2005.  The physical scale is $\simeq 1.1\rm\ kpc\ arcsec^{-1}$.
A cocoon shock appears as an X-ray edge surrounding the radio lobes and
hot spots, which extends from $\simeq30$ arcsec north of the AGN at
the centre to $\simeq60$ arcsec to the west.  There is a great deal of
internal structure, much of which is due to thermal emission from gas
within the shock.  In contrast to the majority of nearby, low power
radio sources at cluster centres, there are no X-ray deficits
over the radio lobes of Cygnus A.  However, comparing the upper and
lower images in Fig.~\ref{fig:cyga}, we see that the X-ray emission is
brighter in regions between the edge of the radio lobes and the cocoon
shocks than it is in adjacent regions over the lobes.  This is as we
should expect.  The gas displaced by the radio lobes must be
compressed into the spaces between the expanding lobes and the shock
fronts, making it brighter than the undisturbed gas and producing a
net excess of X-ray emission over the radio cocoon.  The excess is
greatest where our sight lines through the layer of compressed gas
outside the radio lobes are longest.  The detailed correspondence
between the edges of the radio lobes and enhanced X-ray emission adds
weight to the argument that there are indeed X-ray cavities
corresponding to the radio lobes of Cygnus A, as in lower power
sources.  In addition, \cite{cbk12} have noted that the X-ray deficit
to the south of the AGN corresponds to emission in the 250 MHz LOFAR
map (\cite{mkv11} and these proceedings), arguing that this is older
radio plasma.  The deep X-ray image shows that this central cavity
extends about half as far to the north as it does to the south.  To
the north and west of the AGN the edge of this cavity is remarkably
sharp.  This puts tight constraints on either the physical
properties of the gas and radio plasma, or on the dynamics of
formation of these cavities (\eg \cite{rmf05}, \cite{ps06}).

\begin{figure}[t]
\begin{center}
\includegraphics[width=0.6\textwidth]{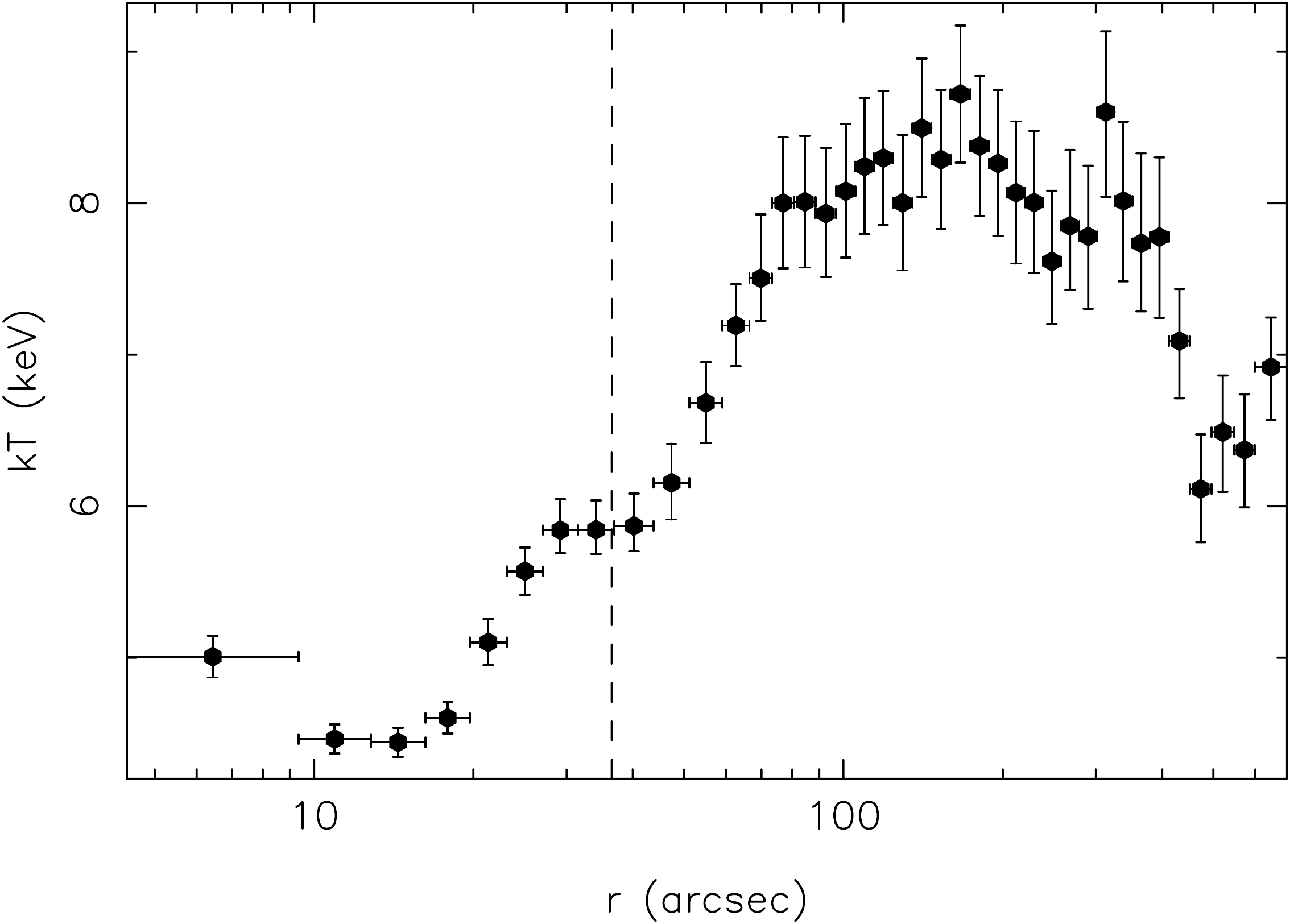}
 \caption{Projected temperature profile of the gas around Cygnus A.
   Temperatures are measured in circular annuli centred on the AGN,
   excluding the regions of bright radio emission.  The dashed
   vertical line marks the radius of the southwest shock front, which
   is smeared by the binning here.  Nevertheless, there is an
   evident bump in the projected temperature $\sim5$ -- 10 arcsec 
   inside the shock, consistent with expectations from fitting the
   surface brightness profile.} \label{fig:tshock} 
\end{center}
\end{figure}

\section{Mean jet power} \label{sec:jetpower}

The cocoon shocks provide an excellent means for estimating the
average power of Cygnus A during its current outburst.  We have fitted
the X-ray surface brightness profile of the shock front in a 50 degree
wedge to the southwest of the AGN with a shock model.  A spherical,
numerical hydrodynamic model was used to compute surface brightness
profiles of shocks due to point explosions at the centre of an
initially isothermal, hydrostatic atmosphere (\eg \cite{nmw05}).  For
the best fitting model, the density in the unperturbed gas varies as
$r^{-1.38}$, the radius of the shock is 40 kpc, its Mach number is
1.37, its age is 16 Myr and its mean power is $\simeq
4\times10^{45}\rm\ erg\ s^{-1}$.  The \chandra{} surface brightness
profile of the shock is very insensitive to the gas temperature, so
that the temperature profile provides an independent test of the
model.  Our assumption that the unshocked gas is isothermal is only
roughly correct.  Nevertheless, the model predicts that the projected
temperature should peak $\simeq5$ arcsec behind the shock, $\simeq
0.6$ keV above the preshock temperature.  Allowing for the preexisting
temperature gradient and smearing due to spherical averaging, the
shock model is consistent with the observed profile of projected
temperature (Fig.~\ref{fig:tshock}).  The spherical model
underestimates the shocked volume, hence the total energy required to
drive the shocks.  A more realistic model would also have continuous
energy injection, which leaves a greater fraction of the injected
energy in cavities, increasing the energy required to produce a given
shock strength.  Thus, we have underestimated
the total outburst energy.  On the other hand, the point explosion
model maximizes the shock speed at early times, causing the shock age
to be underestimated and increasing the power estimate.  The net
result is probably to underestimate the average jet power, but not by
more than a factor of about two.

\section{Particle acceleration in the shock}

\cite{ckh09} found that X-ray emission from the inner portions of the
shock surrounding the southwest lobe of Centaurus A is thermal, while
that from more remote parts of the shock is nonthermal.  The thermal
emission fades out before the nonthermal emission appears, creating a
gap in the X-ray shock front.  There is a similar gap in the Cygnus A
shock front, seen most clearly in the region 15 -- 20 arcsec inside
the western hotspots in Fig. \ref{fig:cyga}.  Is this also due to a
changeover from thermal to nonthermal X-ray emission?  The existing
X-ray data are unable to distinguish between thermal and nonthermal
models for the shocks near the hotspots.  However, it is noteworthy
that the 6 cm radio emission, which is confined within the layer of
shock compressed gas in the inner regions of the radio lobe, extends
right out to the X-ray shock beyond the gap, in the regions around the
hotspots.  This radio emission must arise in the shocked intracluster
medium, rather than within the radio lobes, providing clear evidence
for particle acceleration in the outer parts of the cocoon shocks.
Since the shock fronts near the hotspots are furthest from the AGN,
they must travel fastest.  Allowing for the shape of the fronts and
projection, we estimate that shocks near the hotspots have Mach
numbers of no more than 3.  While the compression is greater in a Mach
3 shock, this is more than offset by the lower preshock gas density and
the smaller radius of curvature of the front near the hotspots.  We
estimate that thermal emission from the shocks near the hotspots
should be no more than $\sim5 \%$ of that at the southwest front.
Thus it appears that the X-ray emission from the shock near the
hotspots is too bright to be thermal.

\section{X-ray jet}

The region of enhanced emission running along the axis of the radio
cocoon, roughly between the two sets of hotspots, is called the X-ray
jet.  
It does not coincide with the radio jet, being considerably wider,
$\simeq 6$ arcsec or $\simeq 6$ kpc, and straighter.  The eastern
X-ray jet is at least as bright as the western X-ray jet, the
approaching side of the radio jet, making it unlikely that Doppler
boosting plays a role in its X-ray emission.  \cite{sbd08} argued that
the X-ray jet is inverse Compton emission from relatively low energy
electrons left behind after passage of the radio jet at earlier times.
An X-ray spectrum extracted from a $11" \times 5.7"$ rectangular
region near its eastern end, with flanking background regions, is
equally well fitted by thermal and nonthermal models, but it is hard
to conceive of a physically reasonable thermal model for the jet
emission.  The power law fit gives a photon index of $1.69\pm0.26$
(90\% confidence).  Assuming this emission is inverse Compton
scattered cosmic microwave background radiation (ICCMB), the
population of electrons required to produce it, with a power law
distribution $dn/d\gamma = A \gamma^{-2.38}$ (to match the slope of
the X-ray spectrum) for Lorentz factors in the range $100 < \gamma <
10000$, alone would have a pressure more than an order of magnitude
larger than the surrounding gas.  Lorentz factors of $\gamma \simeq
1000$ are required to produce 1 keV X-ray photons by scattering the
CMB, so it is difficult to get a lower pressure from a realistic
model.  Magnetic fields and cosmic rays would add to the total
pressure, so an ICCMB model is largely ruled out.  Optical and
ultraviolet radiation from the active nucleus of Cygnus A, or radio
photons from the hotspots might also provide seed photons for inverse
Compton scattering, but in either case we would expect the jet to
brighten significantly towards the photon source, which is not
observed.

Alternatively, the X-ray jet could be due to synchrotron emission.
For a magnetic field strength of $55\ \mu\rm G$, roughly the
equipartition field in the lobes, electrons with $\gamma \simeq
4\times10^7$ are required to produce 1 keV synchrotron photons.  For a
power law electron distribution like that of the ICCMB model, the
required pressure would be $\simeq10^{-5}$ smaller.  However, the
synchrotron lifetimes of the electrons would be only $\simeq 200$ yr,
so a synchrotron model requires \textit{in situ} electron acceleration
(the smallest distance resolvable by \chandra{} in Cygnus A is $\simeq
3000$ light years).  This is much like the Centaurus A jet, but there
the case for X-ray synchrotron emission is strong (\eg \cite{ghc10}).
The main issue for Cygnus A is that an active power source is required
to maintain the population of highly relativistic electrons.  This
problem would be solved if \textit{the X-ray jet reveals the actual
  path of energy flow from the AGN to the hotspots.}  Further
supporting this, the width of the X-ray jet is similar to that of the
brighter hotspots, which would naturally explain their sizes.  The
hotspots are highly overpressured and likely dominated by
relativistic plasma.  To keep a hotspot confined by the ``dentist
drill'' effect, a narrow jet must wander over the whole inward facing
surface of the hotspot in a time that is small compared to the sound
crossing time of the hotspot.  This is incredibly challenging.  It
suggests that we should observe the path of a narrow jet to wobble
side-to-side, on a scale comparable to the width of the hotspot, over
distances along the jet (times) of comparable dimension (or smaller),
but this is not seen.  Of course, if the X-ray jet is the main path of
energy flow, the puzzle is then the origin of the radio jets.


\section{A jet flow model} \label{sec:flow}

\begin{figure}[t]
\centerline{%
\includegraphics[width=0.48\textwidth]{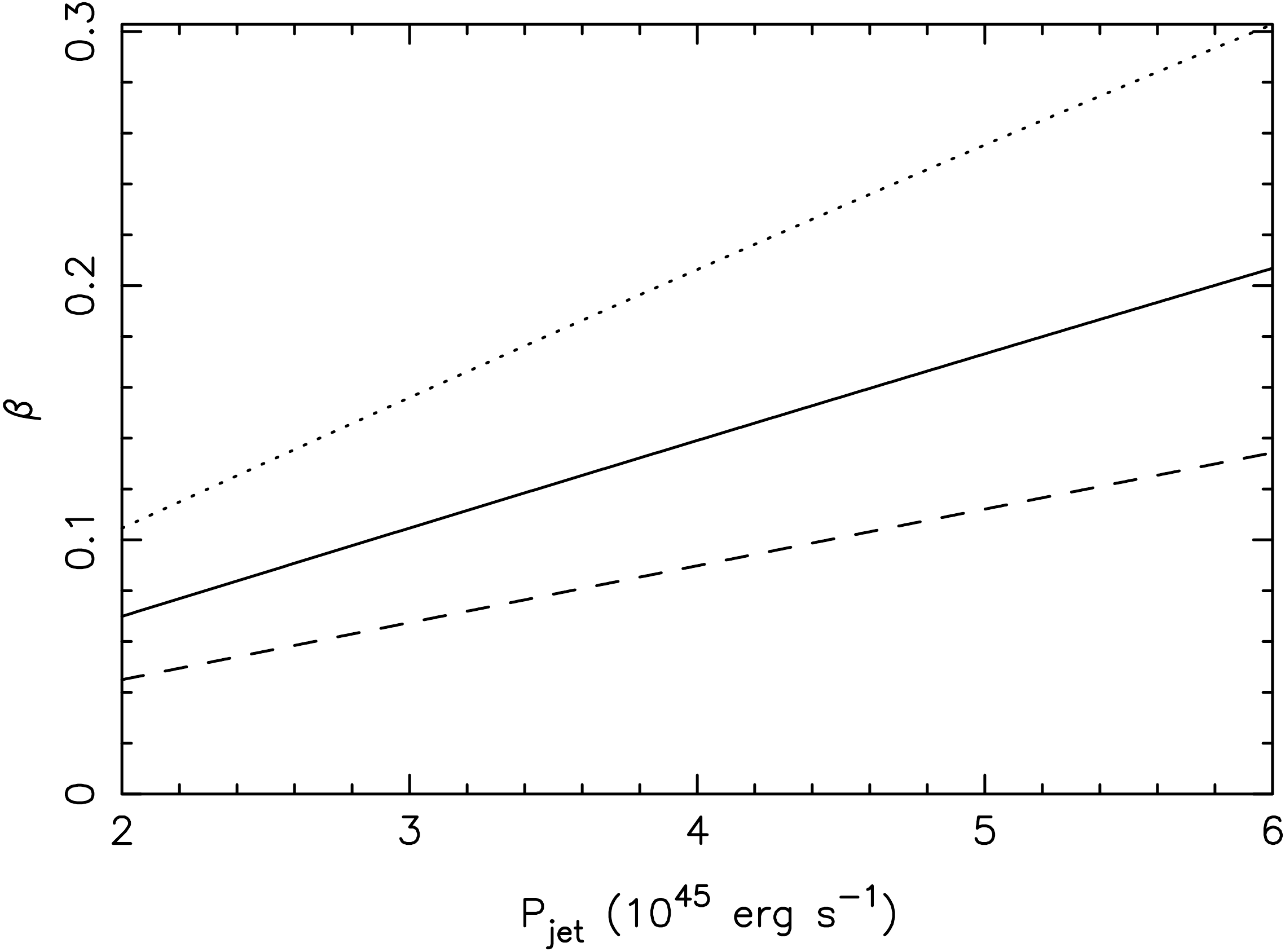}
\includegraphics[width=0.48\textwidth]{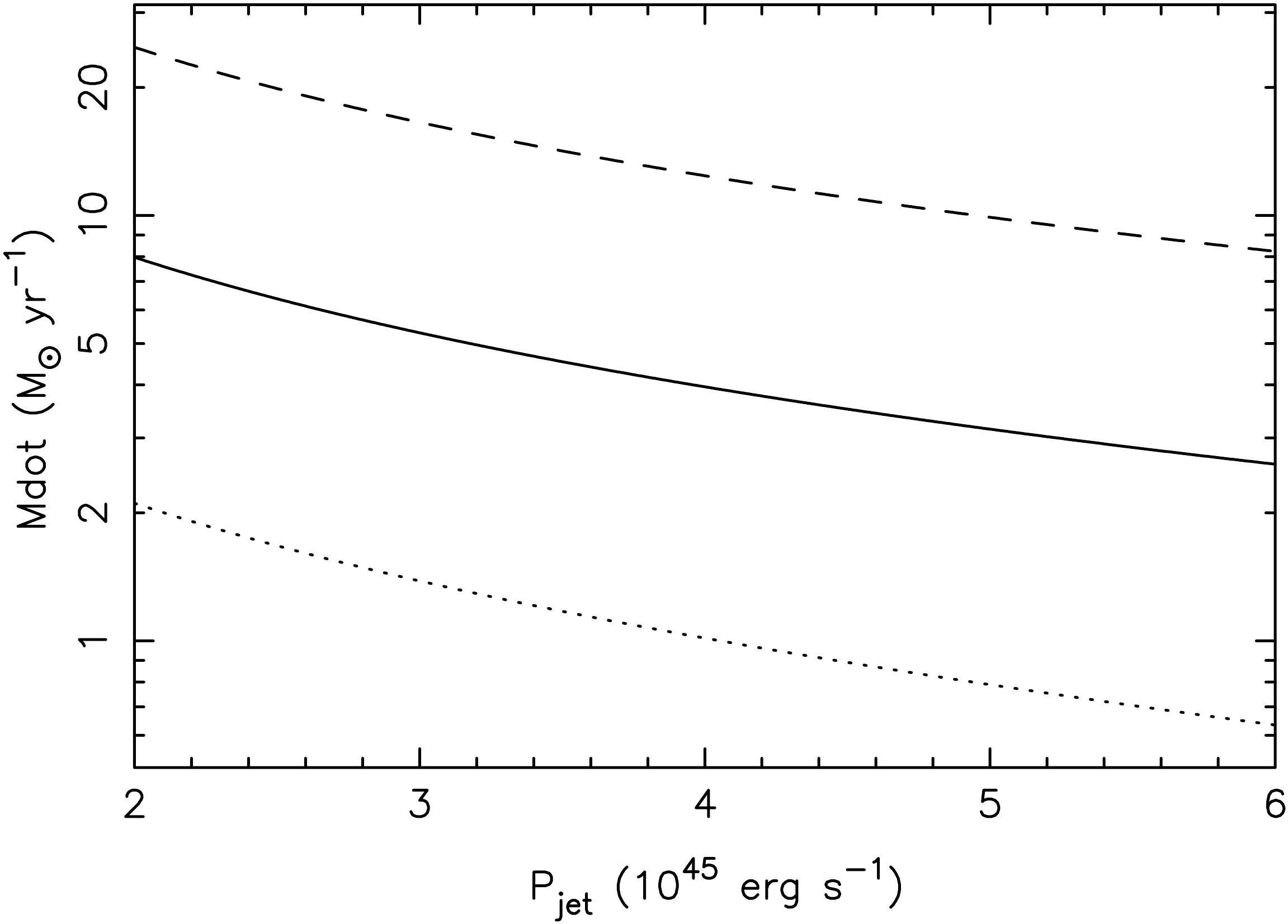}}
\begin{center}
 \caption{Jet flow solutions from section \ref{sec:flow}.  The left
   panel shows $\beta = v/c$ and the right panel shows the flow rate
   of rest mass, both \textit{versus} jet power.  The dotted, full and
   dashed lines are for $\pram/p = 4$, 10 and 20, respectively.}
 \label{fig:flow}
\end{center}
\end{figure}

We use a steady, one-dimensional flow model, like that of \cite{lb02},
to estimate flow parameters for Cygnus A.  The flow rate of rest mass
through the jet is 
\begin{equation} \label{eqn:mass}
\dot M = \rho A c \beta \gamma,
\end{equation}
where $\beta = v/c$, $\gamma$ is the corresponding Lorentz factor,
$\rho$ is the proper density of rest mass in the jet and $A$ is its
cross-sectional area.  The jet power is
\begin{equation} \label{eqn:power}
P = (\gamma - 1) \dot M c^2 + h A c \beta \gamma^2,
\end{equation}
where the enthalpy per unit volume is $h$ and we will assume 
$h = \Gamma p / (\Gamma - 1)$, where $p$ is the pressure and $\Gamma$
is constant.  Lastly, the jet momentum flux is
\begin{equation} \label{eqn:momentum}
\Pi = (P / c + \dot M c) \beta.
\end{equation}
The pressure in a hotspot must be close to the ram pressure of the
jet that runs into it, so we assume $\Pi = A \pram$.  Eliminating
$\dot M$ between the equations 
(\ref{eqn:power}) 
and (\ref{eqn:momentum}) gives
\begin{equation}
{P \over p A c} = \left( {\pram / p \over \gamma + 1} + {\Gamma \over
    \Gamma - 1} \right) \beta \gamma,
\end{equation}
which determines the jet speed, since we have estimates for all the
other quantities.  Synchrotron-self-Compton models give hotspot
magnetic field strengths in the range 150 -- 250 $\mu\rm G$
(\cite{hcp94, wys00}), so that, if the total pressure scales with the
magnetic pressure, $\pram / p \simeq 10$ -- 20.  Using the shock
speeds and external pressures gives lower values, closer to $\pram/p
\simeq 4$.  We assume that the jet is cylindrical with a radius of 3
kpc and a pressure of $p = 3\times10^{-10}\rm\ erg\ cm^{-3}$.  From
section \ref{sec:jetpower}, the power of a single jet is $\simeq
2\times10^{45}\rm\ erg\ s^{-1}$ or somewhat larger.
Fig.~\ref{fig:flow} shows solutions for $\beta$ and $\dot M$ vs jet
power over its plausible range and for $\pram/p = 4$, 10 and 20.  For
reasonable flow solutions, the velocity in the outer part of the jet
is non-relativistic, while the mass flux probably exceeds
$1\rm\ M_\odot\ yr^{-1}$.  It seems likely that most of this mass is
entrained by the jet, rather than flowing all the way from the AGN.

\section{Conclusions}

The radio lobe cavities and cocoon shocks of Cygnus A have much in
common with lower power radio sources in galaxy clusters.
The mean power of the current outburst in
Cygnus A is $4\times10^{45}\rm\ erg\ s^{-1}$, or somewhat larger.
Near to its hotspots, the cocoon shocks of Cygnus A accelerate
electrons, at least to energies sufficient to produce 6 cm radio
synchrotron emission and quite possibly also keV X-rays.  The X-ray
jet of Cygnus A is best explained as synchrotron emission, which
suggests that the X-ray jet, rather than the radio jet, is the main
route of power from the AGN to the hotspots.  Under this assumption,
we find that the outer parts of the jets have flow speeds $v/c \sim
0.1$ and the mass flow rate through the jets is significant.

\acknowledgment

PEJN was partly supported by NASA contract NAS8-03060.

\newcommand\aapr{\textit{ARAA}}
\newcommand\mnras{\textit{MNRAS}}
\newcommand\apj{\textit{ApJ}}
\newcommand\apjl{\textit{ApJ}}
\newcommand\aap{\textit{A\&A}}
\newcommand\nat{\textit{Nature}}

\end{document}